\documentclass[twocolumn,nofootinbib]{revtex4}
\usepackage[dvips]{graphicx}
\usepackage[dvips]{epsfig}
\usepackage{epsf}
\usepackage{epstopdf}
\usepackage{amsmath,bm,epsfig}
\usepackage{epstopdf}
\usepackage{psfrag,color}
\usepackage{multirow}

\begin{document}
\title{Understanding Stokes-Einstein Relation in Supercooled Liquids using
Random Pinning} 

\author{Bhanu Prasad Bhowmik$^{*}$, Rajsekhar Das}
\thanks{These authors contributed equally}
\author{Smarajit Karmakar}
\email{smarajit@tifrh.res.in}

\affiliation{Centre for Interdisciplinary Sciences, Tata Institute 
of Fundamental Research, 21 Brundavan Colony, Narsingi, Hyderabad,
500075, India}

\begin{abstract} 
Breakdown of Stokes-Einstein relation in supercooled liquids is believed to 
be one of the hallmarks of glass transition. The phenomena is studied in 
depth over many years to understand the microscopic mechanism without much
success. Recently it was found that violation of Stokes-Einstein relation 
in supercooled liquids can be tuned very systematically by pinning randomly
a set of particles in their equilibrium positions. This observation suggested
a possible framework where breakdown of Stokes-Einstein relation in the dynamics 
of supercooled liquids can be studied with precise control. We have done 
extensive molecular dynamics simulations to understand this phenomena  
by analyzing the 
structure of appropriately defined set of dynamically slow and fast 
particles clusters. We have shown that the Stokes-Einstein breakdown actually 
become predominant once the cluster formed by the slow particles percolate 
the entire system size. Finally we proposed a possible close connection 
between fractal dimensions of these clusters and the exponents associated 
with the fractional Stokes-Einstein relation.
\end{abstract} 
\maketitle

\section{Introduction}
\begin{figure}[!h]
\begin{center}
\vskip +0.1cm
\includegraphics[scale = 0.34]{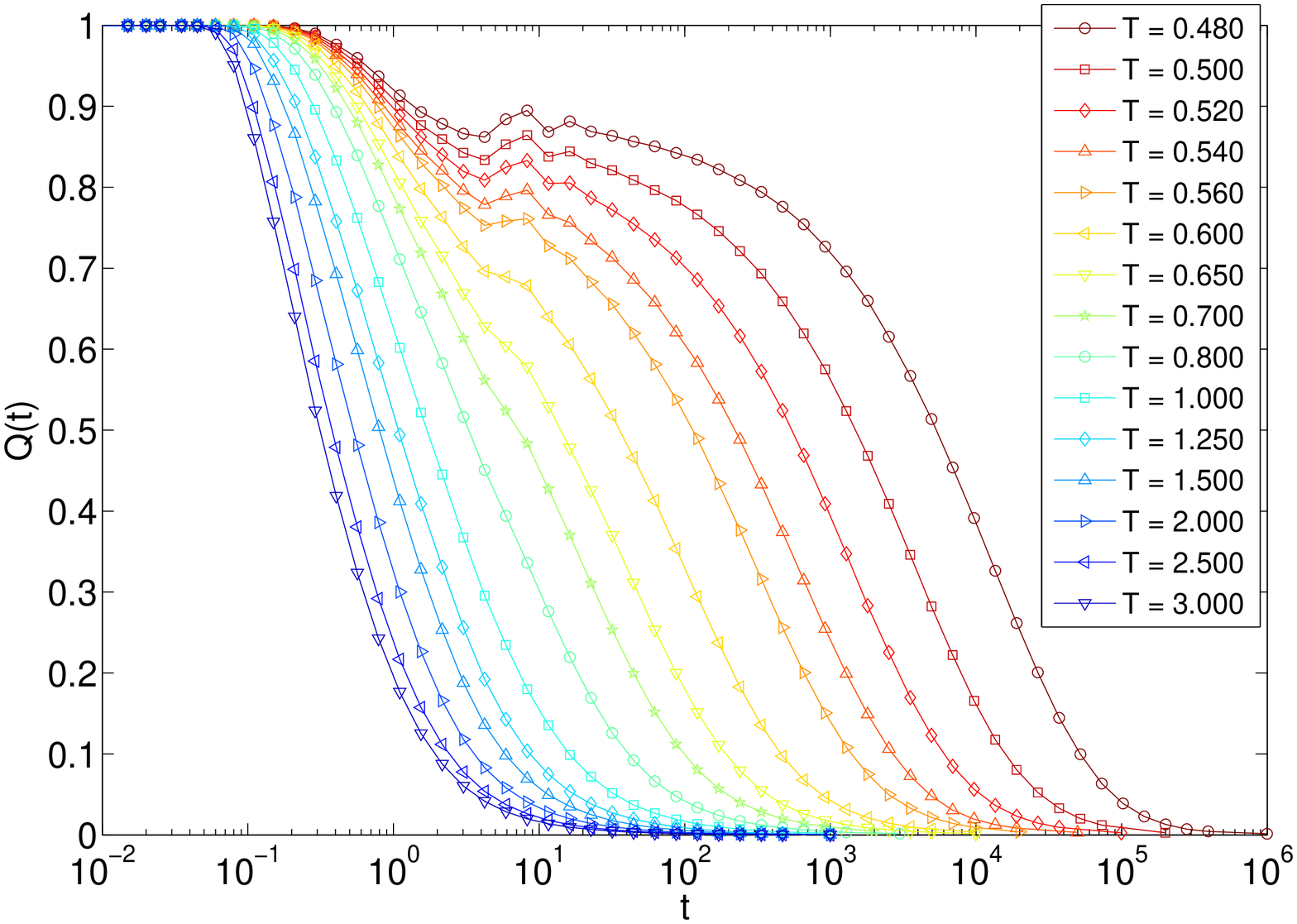}
\caption{Two point density-density correlation function $Q(t)$ (see text for
definition) obtained from molecular dynamics simulations
of a binary glass forming liquids for different temperatures as indicated 
in the legend. Notice the appearance of two step relaxation with decreasing 
temperature.}
\label{qtSketch}
\end{center}
\end{figure}

The dynamics of supercooled liquids and associated glass transition where 
viscosity or relaxation time increases dramatically with decreasing temperature
still puzzle the scientific community even after decades of research 
\cite{spdasRMP,rev1,11BBRMP, rfotrev, kinetic, kinetic2, arcmp,ropp2015}. The 
viscosity change is so dramatic that with few tens of degrees of change in 
temperature, viscosity can change as much as $16$ orders of magnitude and 
the temperature at which viscosity reaches $10^{16}$ Poise is termed the 
calorimetric glass transition temperature, $T_g$. There are many
interesting features seen in the dynamics of supercooled liquids approaching
this calorimetric glass transition that are not observed in normal liquids. 
One such example is the temporal density-density correlation functions which 
in the supercooled temperature regime show multiple relaxation steps.  The 
short time relaxation is termed as $\beta$-relaxation \cite{16KDS} and the 
longer time scale relaxation is called $\alpha$-relaxation as shown in Fig.
\ref{qtSketch}. 
One of the most puzzling phenomena observed in the long time $\alpha$-
relaxation process is the breakdown of the famous Stokes-Einstein relation 
\cite{SER08HM,SER05Einstein,SER87LL}. This is one of the 
most studied phenomena in the context of supercooled liquids without a 
clear general consensus\cite{Th93HS,SEB94KA,SEB95TK,
SEB97CS,SEB00EdigerReview,SEB01MFCR,SEB01ML,SEB06Chen,SEB06BPS,SEB09Xu,SEB10M,
pap:LaNave-etal,13SKDS}. Thus a better understanding of this phenomena will 
definitely be of immediate importance for developing theory of glass
transition.

Stokes-Einstein relation connects viscosity of a liquid ($\eta$) to the 
diffusion constant ($D$) as 
\begin{equation}
D \propto \frac{k_BT}{\eta}
\end{equation}
where $k_B$ is the Boltzmann constant. The constant of proportionality 
depends on the details of the probe particles used to measure the 
diffusion constant of the liquids. One often uses self-diffusion constants 
instead of tracer diffusion in the above equation for simple liquids interacting 
via isotropic pairwise potential. It is found that at least in the 
low temperature regime viscosity, $\eta$ is proportional to the long time
$\alpha$-relaxation time, $\tau_{\alpha}$ \cite{shiJCP2013}
(see method section for definition), so it is customary in the literature 
to cast the Stokes-Einstein relation in the following form
\begin{equation}
D \propto \frac{k_BT}{\tau_{\alpha}}
\end{equation}
where $\eta$ is replaced by $\tau_{\alpha}$. The other reason for this 
modification lies in the fact that calculation of $\tau_{\alpha}$ is much 
easier and straight forward than the viscosity (Notice that it 
is already shown in the literature that this is not the reason for breakdown
in Stokes-Einstein relation). Thus if one plots $D\tau_{\alpha}/T$ as a 
function of $T$, then it should be temperature independent constant, but it 
turns out that for supercooled liquids this ratio becomes a strong function of
temperature below some temperature. Phenomenological arguments \cite{Th93HS, 
SEB95TK} often assume existence of dynamic heterogeneity 
\cite{edigerhet,onuki,harrowell,donati,DH-direct,09KDS,10KDSa, 06KSD, 07CBK} 
which means existence of mobile and less mobile dynamical regions in the 
supercooled liquid states in order to explain the violation of Stokes-Einstein 
relation. Many previous studies indeed showed positive correlation between
Stokes-Einstein breakdown and Dynamic Heterogeneity \cite{06KSD,13SKDS}
as measured by four-point density-density correlation function.

Sometimes a generalized version of the Stokes-Einstein relation 
\cite{SEB06BPS,pap:SE-Douglas-Leporini} is also used to 
describe the inter-relationship between diffusion constant and relaxation 
time in the supercooled regime with the following form 
\begin{equation}
D \propto \tau_{\alpha}^{-1+\omega}
\end{equation}
where exponent $\omega$ characterizes the degree of deviation from normal 
Stokes-Einstein relation. This relation is called fractional Stokes-Einstein
relation and in \cite{SEB06BPS,pap:SE-Douglas-Leporini}, it was shown that 
fractional Stokes-Einstein
exponent can depend on the dimensionality of the system as well as on the 
details of the microscopic interaction between particles. 

From the Mean Field arguments, it is expected that $\omega$ will go to 
zero as one increases the spatial dimensions. This expectation is broadly 
supported by 
the results reported in \cite{DIM09ER,DIM10CIMM,DIM12CPZ,13SKDS}.  A detailed 
numerical study on this aspect \cite{DIM12CPZ} in a hard sphere glass forming 
liquids showed that the fractional Stokes-Einstein exponent $\omega$ actually
goes to zero at eight dimensions. This provided 
for the first time a direct estimate of the upper critical dimension for glass 
forming liquids, above which mean field type description of the liquids should 
hold \cite{DIM12CPZ}. One should keep in mind that computational cost increases 
very rapidly with increasing dimensionality of a system and simulation of any 
mean field model is actually a very difficult numerical task 
\cite{DIM09ER,DIM12CPZ, 13SKDS}. Thus it is needless to say that it would be 
extremely useful if one can change fractional Stokes-Einstein exponent with 
ease in a model system just by tuning some parameters. 

In \cite{14SK}, it was shown that one can change fractional Stokes-Einstein
exponent with ease in model systems simply by randomly pinning some fraction 
of particles in their equilibrium positions \cite{03Kim,12KLP, cammarotaPinning, 
12BK, 13KB, 11KMS, krakoviackPinning, szamelPinning, 15CKD, 15CKDPNAS}. 
Here we will use this protocol
to understand in depth the breakdown of Stokes-Einstein relation and its 
relation with any underlying structural changes in the liquids. The rest 
of the paper is organized as follows: first we will give details of the model
system studied, then will give the definitions of the different 
dynamical quantities calculated in this study with the details of the pinning
protocol used to generate random pinning in the system. Then we will discuss 
the results obtained and finally discuss implications of these results on 
our understanding of breakdown of the Stokes-Einstein relation.

\section{Method and Simulation Details}
\noindent{\bf Model: }We perform simulations for a model glass forming liquids 
in two dimensions 
in this work. The model is characterized by a repulsive inverse power-law 
potential (2dR10) \cite{12KLP}. This model has been studied extensively 
and found to be a very good glass former. The interaction potential is given by
\begin{equation}
V_{\alpha\beta}(r)=\epsilon_{\alpha\beta}
\left(\frac{\sigma_{\alpha\beta}}{r}\right)^{n}
\end{equation}
where $\alpha,\beta \in \{A,B\}$ and $\epsilon_{\alpha\beta}=1.0$, 
for all $\alpha$ and $\beta$. $\sigma_{AA}=1.0$, $\sigma_{AB}=1.18$, 
$\sigma_{BB}=1.40$, and $n = 10$. 
The interaction potential was cut off at 1.38$\sigma_{\alpha\beta}$. For 
this model the system sizes studies are $N = 1000$ and $N = 10000$ at a 
number density $\rho = 0.850$ for many different temperatures in the 
range $T\in[0.480,3.000]$. These simulations are done in the canonical 
ensemble using a modified leap-frog integration scheme with the Berendsen 
thermostat. We have also performed simulation with another constant 
temperature simulation algorithm due to Brown and Clark \cite{brownClark}. 
The results do not depend on the exact algorithm used for integrating the 
equations of motion. We have averaged the data over $32$ independent runs of 
length $100 ~\tau$. 

\vskip +0.5cm
\noindent{\bf Pinning Protocol: }
In this study we have pinned some fraction $\rho$ of total number of particles
randomly from their equilibrium configurations. First we equilibrate
a system of given number of particles, $N$ at the studied temperatures and 
then we choose randomly $N_{p} = \rho N$ number of particles from one such 
equilibrated configuration and pinned their positions. The advantage of such 
pinning protocol is that the system with the pinned particles are already 
in equilibrium as the positions of the pinned particles are taken from an 
equilibrium configurations. This procedure bypasses re-equilibration of the
system after introducing these form of quenched disorder. We then do averaging
over the different realizations of these pinning positions (averaging over 
quenched disorder) to calculate different thermodynamic and dynamic quantities. 
In this study we have done averaging of $32$ different realizations of disorder
and we checked that this amount of averaging is enough to get reliable 
results. One can also pin particles in an ordered arrangements with further
equilibration and results may depend on the details of the pinning protocol 
but we don't expect much qualitative change in the reported results.

\vskip +0.5cm
\noindent{\bf Correlation Functions: }
Dynamic heterogeneity in the system is characterized by the four-point density
correlation function $g_4 (r,t)$ \cite{chandan,silvio,sharon} defined as
\begin{eqnarray}
g_4({\bf r},t) &=& \left[\langle \delta \rho(0,0) \delta \rho(0,t)
\delta \rho({\bf r},0)
\delta \rho({\bf r},t)\rangle\right], \nonumber \\
&-&\left[\langle \delta \rho(0,0)\delta \rho(0,t)\rangle\right]
\left[\langle \delta \rho({\bf r},0)\delta \rho({\bf r},t)\rangle\right]
\label{fourpt}
\end{eqnarray}
where $\delta \rho({\bf r},t)$ represents the deviation of the local density
at point ${\bf r}$ at time $t$ from its average value, and $\langle \cdots 
\rangle$ represents a thermal or time average and $\left[\cdots\right]$ 
represents quenched disorder average. This function quantifies 
the correlation of the relaxation of density fluctuations at two points 
separated by distance $r$. A four-point susceptibility may be defined as the 
$k=0$ value of the Fourier transform $g_4({\bf k},t)$ of this function. A 
variant~\cite{lacevic} of this four-point function has been used extensively 
in numerical studies of dynamic heterogeneity because it is easier to
compute. This quantity is obtained from the overlap function $q(t)$
defined as
\begin{eqnarray}
q(t)= \frac{1}{N-N_p}\sum_{i=1}^{N-N_p}w(|{\bf{r}}_i(0)-{\bf{r}}_i(t)|),
\label{overlap} 
\end{eqnarray} 
where ${\bf{r}}_i(t)$ is the position of particle $i$ at time $t$, $N$ is 
the total number of particles and $N_p$ is the number of pinned particles.
$w(r) = 1$ if $r \le a_0$ and zero otherwise, 
and $a_0$ is a short-distance cutoff chosen to be close to the distance at 
which the root-mean-square displacement of the particles as a function of 
time exhibits a plateau (the precise choice of the value of $a_0$ turns out 
to be qualitatively unimportant). This function may be viewed as the ``self'' 
part of the density correlation function
\begin{equation}
C(t) = \int d{\bf{r}} \rho({\bf{r}},0)\rho({\bf{r}},t),
\label{density_corr}
\end{equation}
with the window function $w(r)$ used to treat particle positions separated by 
distances smaller than $a_0$, due to small-amplitude vibrational motion in the 
"cage'' formed by neighboring particles, to be the same. The thermal averaged
and disorder averaged overlap function 
$Q(t) = \left[\langle q(t)\rangle\right]$. The fluctuations in this two point function yields the dynamical four-point 
susceptibility: 
\begin{equation}
\chi_4(t)=\frac{1}{N-N_p}\left(\left[\langle q^2(t)\rangle\right] -\left[\langle q(t)\rangle\right]^2\right).
\label{chi4}
\end{equation}
with the peak value defined as $\chi_4^P \equiv \chi_4(t = \tau_4)$, 
where $\tau_4$ is the time at which $\chi_4(t)$ attains its maximum value 
and $\tau_4 \simeq \tau_{\alpha}$ at all temperatures.
The $\alpha$-relaxation time $\tau_{\alpha}$ is defined using the two-point 
correlation function as $Q(t = \tau_{\alpha}) = 1/e$.

\vskip +0.5cm
\noindent{\bf Diffusion Constant: }
The diffusion constant $D$ for different temperatures and pinning density 
is calculated from the slope of the mean squared displacement using the 
relation 
\begin{equation}
\langle \Delta r^2(t)\rangle = 2dDt
\end{equation}
where $d$ is the spatial dimension and in this case $d = 2$.

\vskip +0.5cm
\noindent{\bf van Hove function: }
van Hove function, $G(x,t)$ is the probability of finding a displacement of
a particle of amount $x$ over a timescale of $t$. It is formally defined 
as follows
\begin{equation}
G(x,t) = \sum_{i,j = 1}^{N}\delta(x - x_i(t) - x_j(0))
\end{equation}
where $x_i(t)$ is one of the coordinates of the position of particle $i$ at 
time $t$ and the self part of this correlation function is defined as 
\begin{equation}
G_s(x,t) = \sum_{i = 1}^{N}\delta(x - x_i(t) - x_i(0))
\end{equation}
Here we have calculated only self part of the van Hove correlation function 
for simplicity. 

\section{Results}
\begin{figure}[!h]
\begin{center}
\hskip -0.2cm
\includegraphics[scale = 0.38]{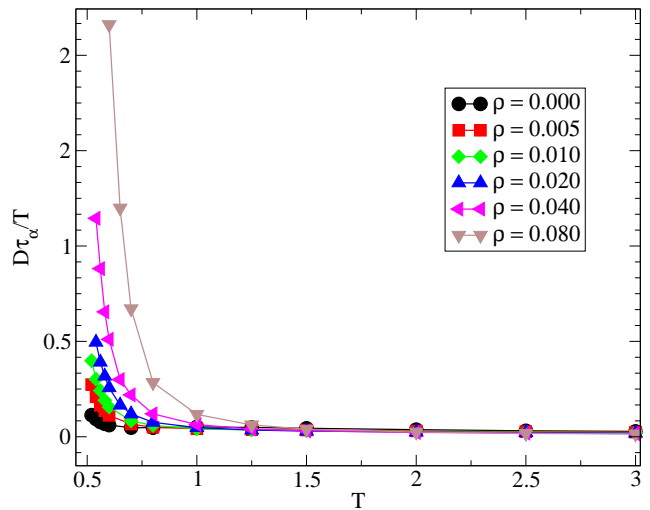}
\caption{The Stokes-Einstein parameter $D\tau_{\alpha}/T$ is plotted as a 
function of temperature for different pinning concentration, $\rho$. One
can clearly see that at high temperatures the parameter is almost independent 
of temperature and starts to become temperature dependent somewhat sharply.
The temperature at which the breakdown happens increases to 
higher value with increasing pinning concentration.}
\label{seViolation}
\end{center}
\end{figure}
In Fig.\ref{seViolation}, we have plotted the Stokes-Einstein parameter
$D\tau_{\alpha}/T$ as a function of temperature for different pinning 
density. One can clearly see that the quantity is fairly independent of 
temperature up to some temperature for a given pinning concentration and 
then it start to increase, indicating the violation of 
Stokes-Einstein relation in these temperature regimes. The phenomena 
becomes even more prominent with increasing pinning concentration and the
Stokes-Einstein parameter becomes strongly temperature dependent and the
temperature at which this violation happens seems to also increase with 
increasing pinning concentration as shown in Fig.\ref{seViolation}. Thus 
pinning a set of randomly chosen particles in their equilibrium positions
will be an excellent model system to understand breakdown of Stokes-Einstein
relation in supercooled liquids. 

\begin{figure}[!h]
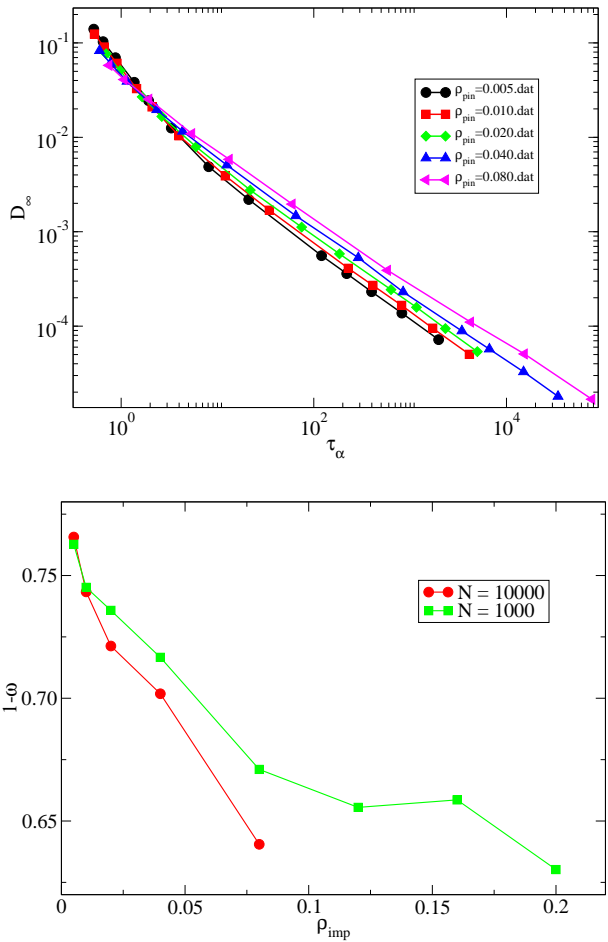

\begin{center}
\vskip +0.1cm
\includegraphics[scale = 0.42]{DvsTau.eps}
\vskip +0.5cm
\includegraphics[scale = 0.350]{oneMinusOmega_10k.eps}
\caption{Top panel: Fractional Stokes-Einstein relation, $\tau_{\alpha}$
is plotted as function of $D$ for different pinning concentrations and
one can see that at higher temperature $D \sim 1/\tau_{\alpha}$, but
at lower temperatures this relation breaks down and one gets 
$D \sim \tau_{\alpha}^{-1+\omega}$ with $\omega > 0$. This exponent
seems to increase with increasing pinning concentration as shown in
the Bottom panel. Note that $\omega$ seems to significant finite size
effect and increases faster with 
increasing pinning concentration for $N = 10000$ system size.}
\label{fractionalSE}
\end{center}
\end{figure}
In the same context it is important to discuss the fractional Stokes-Einstein
relation and how the exponent depends on the pinning density. In 
Fig.\ref{fractionalSE}, we have plotted $D$ vs $\tau_{\alpha}$ in log-log plot 
for different pinning density. As in \cite{14SK}, we have found that at higher 
temperatures $D \propto \tau_{\alpha}^{-1}$, but as one goes to lower 
temperature, this relation breaks down and one gets 
$D \propto \tau_{\alpha}^{-1+\omega}$, with $\omega > 0$. The value of the
fractional Stokes-Einstein exponent $\omega$ increases continuously with 
increasing pinning density as shown in lower panel of 
Fig.\ref{fractionalSE}. These observations also give us an opportunity 
to gain better understanding of the phenomena without much computational 
difficulty. For the rest of the paper we will discuss about the correlation 
between different dynamical measures of Dynamic Heterogeneity and their
correlation with the breakdown of Stokes-Einstein relation.   
\begin{figure}[!h]
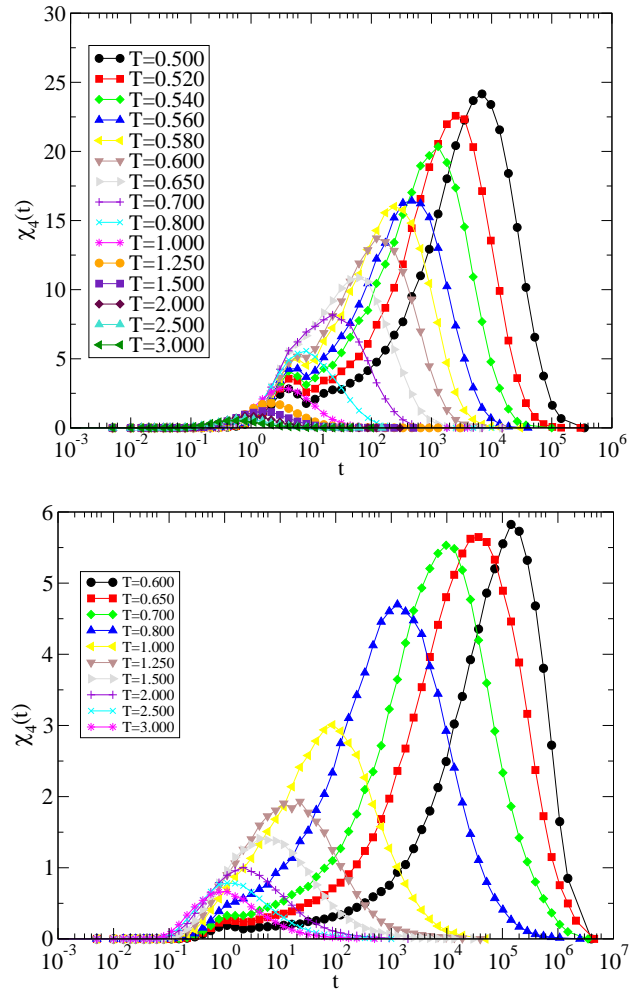

\begin{center}
\includegraphics[scale = 0.37]{{Xi4_0.010}.eps}
\vskip +0.3cm
\includegraphics[scale = 0.36]{{Xi4_0.120}.eps}
\caption{Top panel: Four-point dynamic susceptibility, $\chi_4(t)$ as a 
function of time for different temperatures. Notice the monotonic increase
in the peak height of $\chi_4(t)$ with decreasing temperature. This is 
for pinning concentration, $\rho = 0.020$. Bottom panel: Similar plot but 
for higher pinning concentration $\rho = 0.120$. Notice although the peak
height increases monotonically the absolute value of this function is many
times smaller than the $\rho = 0.020$ case.}
\label{x4vst}
\end{center}
\end{figure}

As discussed before many phenomenological theories suggest Stokes-Einstein breakdown to be due to the presence of dynamic heterogeneity in the system
at the supercooled temperature regime \cite{SEB95TK,13SKDS} and indeed some
correlation was also found in \cite{13SKDS}. It was found that peak height of
$\chi_4$ starts to grow rapidly at a temperature very close to the temperature
where Stokes-Einstein breakdown seems to happen for different model system 
in different dimensions. To see whether similar correlation exists here for 
the pinned system, we have calculated the four-point susceptibility 
as shown in different panels of Fig.\ref{x4vst}. One can see the usual 
temperature dependence of $\chi_4(t)$ for different pinning density $\rho$.

\begin{figure}[!h]
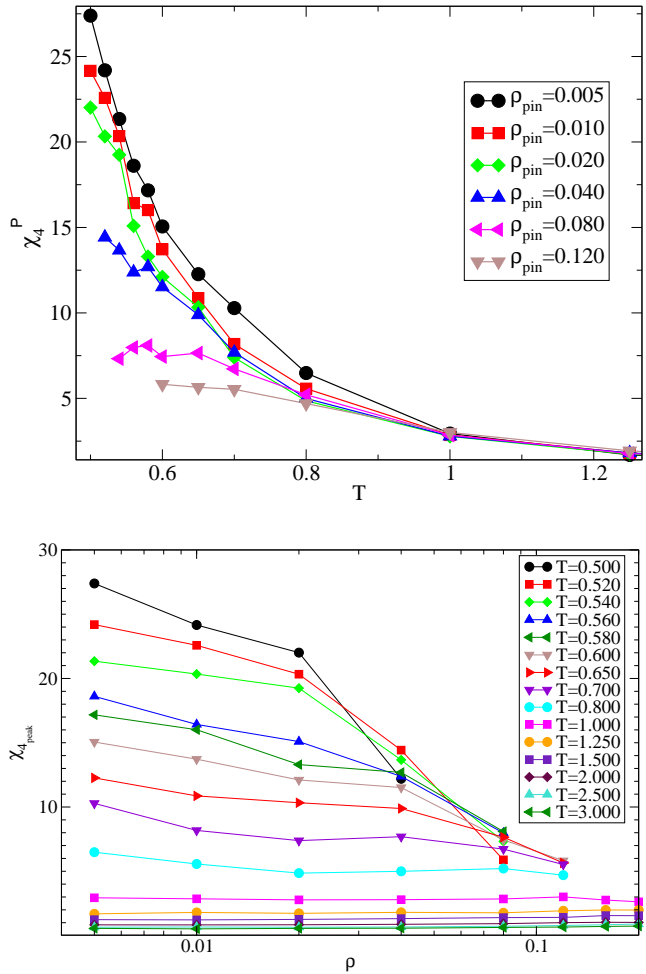

\begin{center}
\vskip +0.1cm
\includegraphics[scale = 0.40]{Xi4PeakVst.eps}
\vskip +0.5cm
\includegraphics[scale = 0.30]{Xi4PeakvsRho.eps}
\caption{Top panel: Peak height of $\chi_4(t)$, $\chi_4^P$ as a function 
of temperature for pinning concentration. Notice that temperature dependence
of $\chi_4^P$ become much more weak with increasing pinning concentration. 
Bottom panel: $\chi_4^P$ as a function of pinning concentration for different
temperatures.}
\label{x4pVstandRho}
\end{center}
\end{figure}

It is interesting to note that actual peak height of $\chi_4(t)$ decreases
with increasing pinning density as can be seen in top panel of 
Fig.\ref{x4pVstandRho}, where temperature dependence of $\chi_4^P$ is plotted
for different pinning density and one sees that $\chi_4^P$ decrease with 
increasing pinning density. It becomes much clearer when $\chi_4^P$ is plotted
as a function of pinning density for different temperature as shown in the 
bottom panel of the same figure.  This clearly shows that $\chi_4^P$ 
completely fails to capture Stokes-Einstein breakdown  for these model 
system. Actually it shows anti-correlation.  

Next we check whether the exponent $\beta$, related to stretched exponential
decay of two-point correlation function is correlated with
the Stokes-Einstein breakdown as found in \cite{13SKDS}. We followed
the similar fitting procedure as described in \cite{13SKDS}, to get the 
exponent $\beta$ by fitting the $Q(t)$ data to the following fitting function
\begin{equation}
Q(t) = A \exp{\left[-(\frac{t}{\tau_1})^2\right]} + B \exp{\left[-(\frac{t}{\tau_\alpha})^{\beta}\right]}
\label{qtFitFn}
\end{equation}
\begin{figure}[!h]
\begin{center}
\vskip +0.1cm
\includegraphics[scale = 0.38]{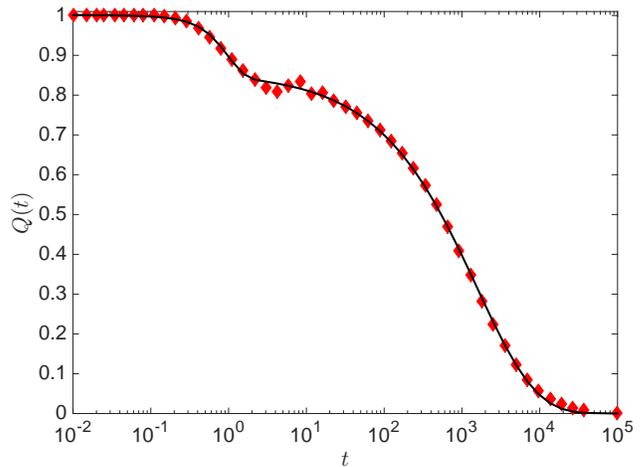}
\caption{Fitting of $Q(t)$ using Eq.\ref{qtFitFn} for temperature $T = 0.520$
and $\rho = 0$. The fitting observed is very good over the whole time window
and thus one can be confident of the obtained values of stretched exponent 
$\beta$.}
\label{qtFitting}
\end{center}
\end{figure}
Although it is a non-linear fitting with $5$ parameters, with good initial 
choice of the parameters the fitting converges to the right solution very 
quickly without any problem as can be seen in Fig.\ref{qtFitting}. The quality
of the fit gives us the confidence on the reliability of the extracted values 
of the exponent $\beta$.

\begin{figure}[!h]
\begin{center}
\vskip +0.1cm
\includegraphics[scale = 0.4]{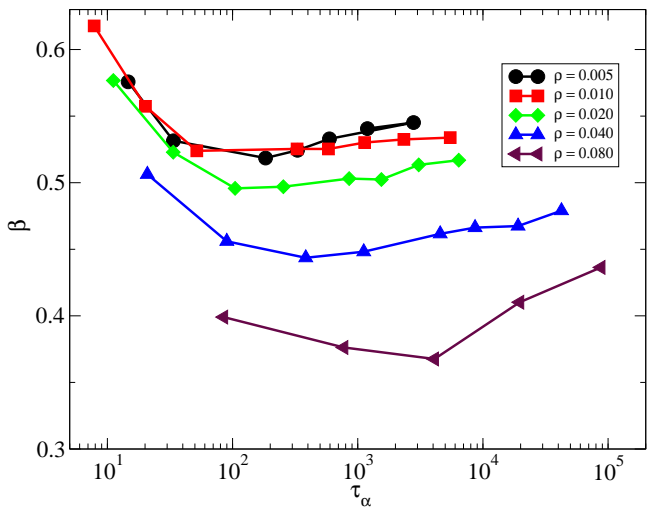}
\caption{Stretched exponent $\beta$ as a function of $\alpha$-relaxation 
time scale $\tau_{\alpha}$ for different pinning concentrations. Notice 
that $\beta$ becomes significantly smaller for larger pinning concentration
at lower temperature.}
\label{betaVsT}
\end{center}
\end{figure}
In Fig.\ref{betaVsT}, we have shown the stretched 
exponent $\beta$ as a function of temperature for different pinning density
and one can clearly see that $\beta$ becomes smaller and smaller with 
increasing pinning density. In \cite{13SKDS}, it was shown that $\beta$ becomes
closer to one as one increases dimensions of the system from $d=2$ to $d=4$
and Stokes-Einstein breakdown also becomes less prominent with increasing
dimensionality. Thus it seems stretched exponent $\beta$ to be an excellent
identifier of the Stokes-Einstein breakdown even for the pinned systems.

Now if we assume that stretched exponential relaxation is related to 
hierarchical relaxation processes, then one can write 
\begin{equation}
Q(t) \sim \exp{\left( -t/\tau_{\alpha}\right)^{\beta}} =  \int_{0}^{\infty}P(\tau)\exp{\left( -t/\tau\right)}d\tau
\end{equation}
where $\beta$ will be related to the relative variance of the distribution and
smaller $\beta$ will corresponds to larger relative width of the distribution.
To ascertain that we have discretized the above equation to calculate the 
distribution $P(\tau)$ by minimizing $\chi^2$,
\begin{equation}
\chi^2 = \sum_{i=1}^{m}\left[ Q(t_i) - \sum_{j=1}^{M}P_j\exp{(-t_i/\tau_j)}
\Delta\tau\right]^2
\label{chiSq}
\end{equation}
\begin{figure}[!h]
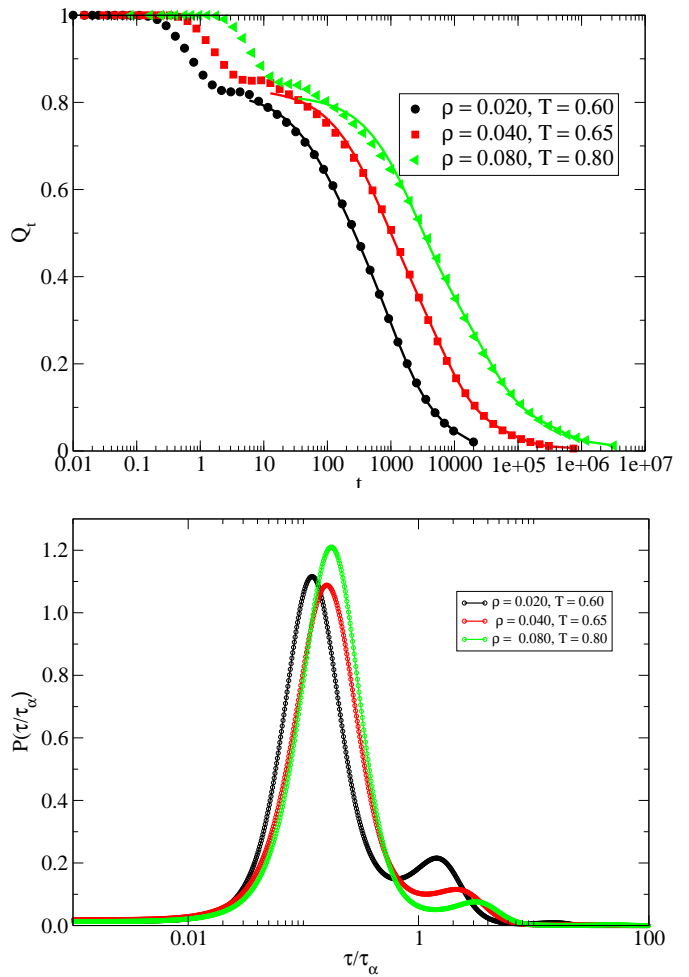

\begin{center}
\vskip +0.1cm
\includegraphics[scale = 0.40]{fitQt.eps}
\vskip +0.3cm
\includegraphics[scale = 0.35]{distrelaxtimeTau.eps}
\caption{Top panel: Convergence of $\chi^2$-optimization, Eq.\ref{chiSq}. Time 
axes of $Q(t)$ for $T=0.65, \rho = 0.040$ and $T=0.80, \rho = 0.080$ are 
multiplied by $3$ and $8$ respectively for the clarity. Bottom panel: The
obtained distribution of relaxation time, $P(\tau)$. The probabilities
for different pinning concentrations and temperatures are plotted after rescaling
the timescale by $\tau_{\alpha}$ for better comparison. Notice that distribution
gets fatter and fatter for larger values of $\tau$ with increasing pinning 
concentration, completely consistent with exponent $\beta$ being smaller in 
the respective cases.}
\label{pTau}
\end{center}
\end{figure}
where $m$ is the number of time points $t_i$, where two point-correlation 
function $Q(t)$ is obtained from simulations and $M$ is the number of 
discretized $\tau$ values ($\tau_j$) where we want to find out the probability
density $P_j$. $\Delta\tau = \tau_{j+1} - \tau_j$, is the elementary 
discretization step size. One crucial transformation which is required to 
get the required convergence of the $\chi^2$ minimization is to do the following
variable change, $P_j = p_j^2$ and then minimize with respect to the new 
variable $p_j$. This ensures the positivity of the probability
\cite{chandanDFT}. As shown in top panel of Fig.\ref{pTau}, 
the convergence is quite good 
and one can get very reliable estimate of the underlying probability 
distribution of the relaxation time. We have calculated the probability for
all the different temperatures at different pinning density but showing the
results for three different pinning concentrations at temperatures where
$\tau_{\alpha}$ are comparable to each other. The results clearly show 
that indeed the 
normalize distribution tends to have fatter tail in the larger $\tau$ values
with increasing pinning density. The nice bi-modality of the distribution 
is also interesting and will discuss this issue in the context of distribution
of diffusion constants.  

Still now we have shown that peak height of $\chi_4(t)$ does not seem to 
capture the increasing degree of Stokes-Einstein violation in the pinned 
system, whereas stretched exponent $\beta$ and the associated distribution
of relaxation time seem to correlate very well. Next we will  
quantify the dynamic heterogeneity by identifying "slow" and "fast" particles
over the $\alpha$-relaxation time scale as done in \cite{14SK} and show
the results on the clustering properties of these slow particles in the 
system to elucidate their role in breakdown of Stokes-Einstein relation.

\begin{figure}[!h]
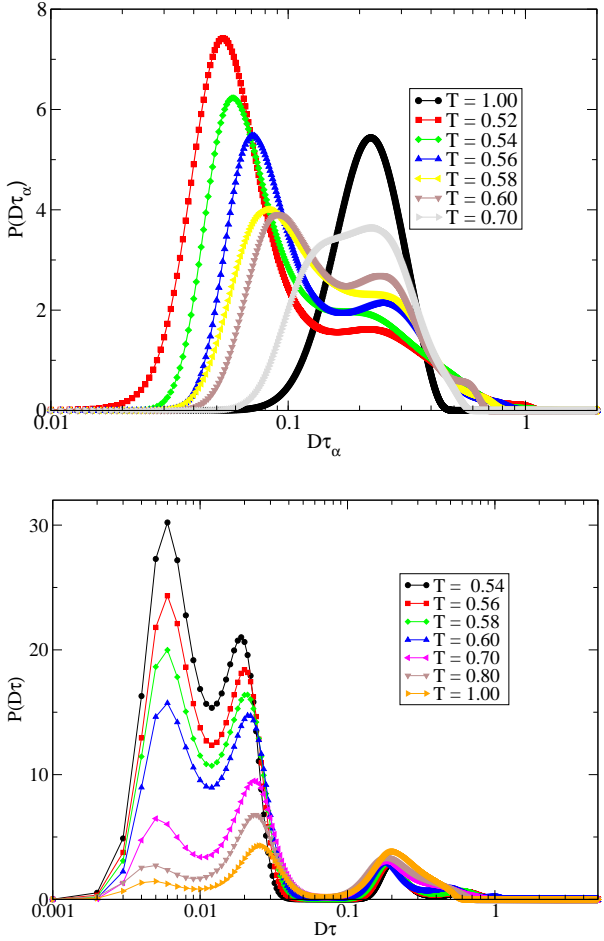

\begin{center}
\includegraphics[scale = 0.37]{{r0.000}.eps}
\vskip +0.5cm
\includegraphics[scale = 0.32]{{r0.020}.eps}
\caption{Distribution of diffusion constants for different temperatures for
$\rho = 0.00$ (top panel) and $\rho = 0.020$ (bottom panel). Notice the 
bimodal distribution at lower temperatures.}
\label{pD}
\end{center}
\end{figure}
We will first briefly discuss the method used to obtain the distribution 
of diffusion constants from the van Hove correlation function using Lucy 
iteration \cite{74Lucy}. This method is also recently used in
\cite{12WKBG} for the diffusion processes in biological systems. We start
with the following assumption: that particles' displacements are mainly
caused by diffusion processes. Due to dynamic heterogeneity in the system
at least in the supercooled temperature regime it is not very surprising to
expect that there will be a  distribution of local diffusivity $p(D)$. 
Then we can express van Hove correlation function calculated at 
$\alpha$-relaxation timescale, 
$G_s(x,\tau_{\alpha})$ in terms of $p(D)$ as
\begin{equation}
G_s(x,\tau_{\alpha}) = \int_0^{D_0} p(D). g(x|D). dD,
\label{vanHoveEq}
\end{equation}
where $g(x|D) = 1/\sqrt{4\pi D\tau_{\alpha}} \exp\left( -x^2/4D\tau_{\alpha} 
\right)$ and $D_0$ 
is the upper limit of diffusion constant and will be equal to diffusivity for a 
free diffusion. Now given the $G_s(x,\tau_{\alpha})$ we calculate the 
distribution of diffusivity $p(D)$ following \cite{74Lucy} as
\begin{equation}
p^{n+1}(D) = p^{n}(D)\int_{-\infty}^{\infty} \frac{G_s(x,\tau_{\alpha})}{G^n_s(x,\tau_{\alpha})}g(x|D) dx,
\end{equation}
where $p^n(D)$ is the estimate of $p(D)$ in the $n^{th}$ iteration with 
$p^0(D) = (1/D_{avg})\exp(-D/D_{avg})$ and
\begin{equation}
G^n_s(x,\tau_{\alpha}) = \int_0^{D_0} p^n(D). g(x|D). dD.
\end{equation}
Similarly
\begin{equation}
P^{n+1}(D\tau_{\alpha}) = P^{n}(D\tau_{\alpha})\int_{-\infty}^{\infty} \frac{G_s(x,\tau_{\alpha})}{G^n_s(x,\tau_{\alpha})}g(x|D) dx,
\label{eqn:LucyDtau}
\end{equation}
where $p(D)dD = P(D\tau_{\alpha})d(D\tau_{\alpha})$. The choice of 
$D\tau_{\alpha}$ as our variable is due 
to the fact that $D$ changes by several orders of magnitude in the studied 
temperature range whereas $D \tau_{\alpha}$ changes relatively modestly with 
decreasing temperature and it will be easier to compare the distribution 
obtained for different temperatures (Please see \cite{14SK} for further details).

In Fig.\ref{pD}, we have shown the obtained distribution of
diffusion constants for different pinning concentrations. For the unpinned 
case the distribution is very nicely bimodal below some temperatures, clearly
telling us that one can easily identify a set of "slow" and "fast" moving 
particles over $\alpha$-relaxation time scale. Notice that at much larger 
time than $\tau_{\alpha}$ the dynamic heterogeneity will be averaged out
and one expects the corresponding van Hove function to become Gaussian. Thus
at longer timescale the distribution will also loose its bimodality, but
as already shown in \cite{14SK}, the bimodal feature remains intact even
for timescale which are at least couple of time larger than the $\alpha$-
relaxation timescale. Thus we believe that the results reported here to be 
qualitatively valid even for timescale larger than the typical relaxation 
time. 

\begin{figure}[!h]
\begin{center}
\vskip +0.5cm
\includegraphics[scale = 0.37]{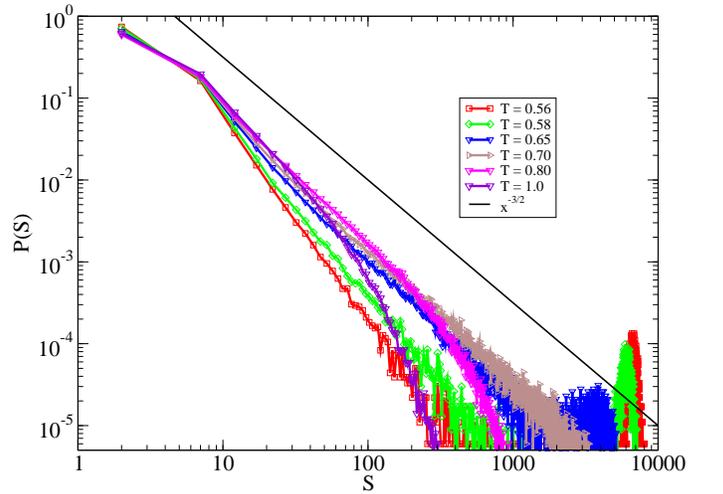}
\caption{Distribution of cluster size, $P(S)$ for different temperatures
for $\rho = 0.040$. Notice the development of nice power law regime at 
lower temperatures with slope $-3/2$. At still lower temperatures one 
can see appearance of a extra hump at larger cluster sizes. This corresponds
to system spanning clusters (see text for details).}
\label{distCluster}
\end{center}
\end{figure}
The cutoff we chose to define
such set of particles is the minimum of the two peaks of the bimodal 
distribution. For system with pinning, the peak associated with the "slow"
particles in the distribution splits into two peaks. This can be understood
as follows: with increasing pinning density the difference in the 
diffusion constants between "slow" and "fast" particles will increase strongly
and there will be a set of particles who reside on the boundary of "slow" 
moving and "fast" moving particles with intermediate diffusion constants. The
split of the first peak in $P(D\tau_{\alpha})$ we believe is due to these 
set of particles which becomes prominent with increasing difference in 
diffusivity between "slow" and "fast" particles. In our studies we have 
considered the particles with these intermediate diffusion constants to be 
part of the "slow" particles. As mentioned before, $P(\tau)$ also shows 
bimodality and it will be interesting to see whether different peaks in that
distribution actually correspond to relaxation of "slow" or "fast" moving 
particles.

Next we have defined a cutoff distance $r_c = \sqrt{4D_c\tau_{\alpha}}$
at the minimum of the two peaks of the $P(D\tau_{\alpha})$ distribution for
a given temperature and pinning density. With this definition we define a 
particles "slow" if it has not moved a distance bigger than the critical 
distance over $\tau_{\alpha}$ timescale and the others who have move beyond
are termed as "fast" particles. In Fig.\ref{distCluster}, 
we have plotted the distribution of cluster size for "slow" moving particles
for different temperature at pinning concentration $\rho = 0.040$. One can
see the appearance of power law distribution at lower temperatures with 
exponent $P(S) \sim S^{-3/2}$. At still further lower temperatures one
can see an appearance of a small hump at larger cluster size, indicating 
the existence of system spanning cluster. Motivated by this observation 
we then have tried to study a possible underlying percolation transition
associated with the clusters of "slow" moving particles.
\begin{figure}[!h]
\begin{center}
\includegraphics[scale = 0.40]{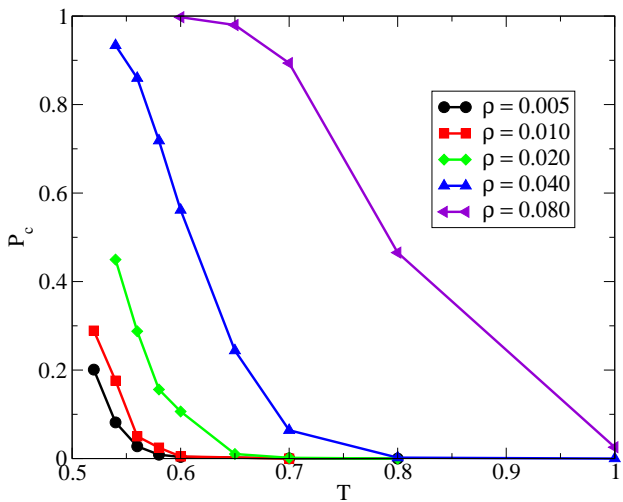}
\caption{Probability of finding system spanning cluster of $N = 10000$ as
a function temperature for different pinning concentrations. Notice the 
sharp change in probability at some critical temperature which changes to 
higher temperatures with increasing pinning concentrations.}
\label{pcVsT}
\end{center}
\end{figure}

\begin{figure}[!h]
\begin{center}
\vskip +0.5cm
\includegraphics[scale = 0.350]{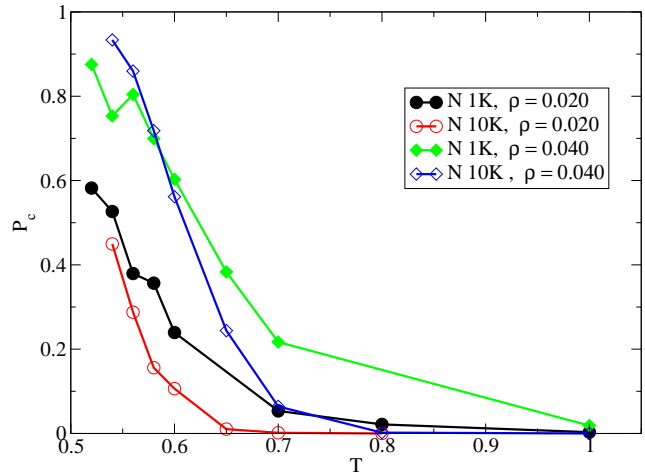}
\caption{Probability of finding system spanning cluster, $P_c$ is plotted 
for two different system sizes $N = 1000$ and $N = 10000$ for two different
pinning concentrations. It is clear that growth of $P_c$ for $N = 10000$ 
system size is much more steeper than $N = 1000$ system, in complete agreement
with an underlying percolation transition.}
\label{fss}
\end{center}
\end{figure}

In Fig.\ref{pcVsT}, we have plotted the probability of finding a system spanning
cluster, $P_c$ as a function of temperature for different pining concentration. 
It is clear that probability of finding system spanning cluster seems to grow 
sharply after a particular critical temperature and this temperature also 
shifts to higher values with increasing pinning concentration. Thus we have a 
clear one to one correspondence between onset of Stokes-Einstein violation 
and the appearance of system spanning cluster. To establish a clear percolation
transition one needs to perform a detailed finite size scaling analysis which 
we have not yet performed here in this study due to large computational 
requirements. 

We also have studied the phenomena for another system size with $N = 1000$ 
particles.
For percolation transition one expects to have a somewhat smoother increase
of probability of finding system spanning cluster with decreasing temperature
due to finite size effects. 
In Fig.\ref{fss}, we have plotted $P_c$ as a function of temperature for two
different system sizes $N = 1000$ and $N = 10000$ for different pinning 
concentrations. One can clearly see that $P_c$ increases somewhat less sharply
with decreasing temperature for smaller system sizes. Thus confirming a 
possible underlying percolation transition \cite{StaufferAharony} associated 
with Stokes-Einstein violation. It will be interesting to perform a detailed 
finite size scaling analysis to calculate the exponents related to this 
transition and its connection to slowing down in dynamics in supercooled 
liquids.   

\begin{figure}[!h]
\begin{center}
\includegraphics[scale = 0.38]{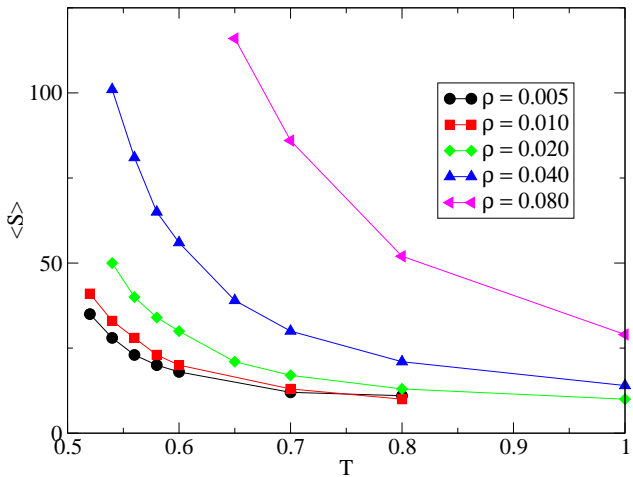}
\caption{Temperature dependence of mean cluster size $\langle S \rangle$ for
different pinning concentrations. $\langle S \rangle$ shows much stronger
temperature dependence with increasing pinning strength, thus confirming 
growth of Dynamic Heterogeneity in the system.}
\label{meanCluster}
\end{center}
\end{figure}
Next we have calculated  the mean cluster size as a function of temperature for 
different pinning concentrations as another indicator of dynamic heterogeneity. 
As shown in Fig.\ref{meanCluster}, the growth of the mean
cluster size is very dramatic and the growth is even more dramatic for 
higher pinning concentrations, suggesting that indeed dynamic heterogeneity
increases with increasing pinning concentration. 

\begin{figure}[!h]
\begin{center}
\hskip -0.5cm
\includegraphics[scale = 0.47]{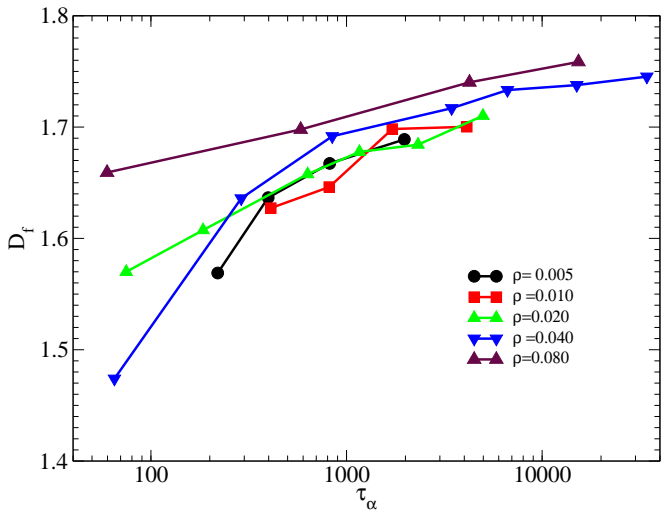}
\caption{Fractal dimension of spanning cluster, $D_f$ as a function of 
temperature for different pinning concentrations. The fractal dimension
is systematically larger for system with larger pinning concentrations
(see text for detailed discussion).}
\label{dfvsT}
\end{center}
\end{figure}
Finally to understand the origin of fractional Stokes-Einstein exponent, 
$\omega$, we have calculated the fractal dimensions of these spanning 
cluster as shown in Fig.\ref{dfvsT}. We have used Sandbox method for 
determining the fractal dimension of these percolating clusters  
\cite{fractalDim}. It is clear that the clusters are not 
compact and the fractal dimensions are in the range $D_f\in[1.5 - 1.8]$ at 
different temperatures and pinning concentrations. The important thing to 
notice is that the fractal dimension is systematically larger for the system 
with increasing pinning concentrations. Thus it seems there is a positive 
correlation between fractal dimensions of the system spanning cluster and the 
fractional Stokes-Einstein exponent.    
Right now we do not have a very clear understanding about this correlation 
but the following simple arguments seem to suggest such correlation. At 
higher temperatures $D \sim 1/\tau_{\alpha}$ and then at lower temperatures
$D \sim 1/\tau_{\alpha}^{1-\omega}$ with $\omega>0$, which implies that 
$\tau_{\alpha}$ increases much faster than the rate at which $D$ decreases.
This difference in change of $\tau_{\alpha}$ and $D$ with decreasing 
temperature will be much more prominent if the cluster forms by the "slow"
moving particles become more compact. This is due to the fact that 
$\tau_{\alpha}$ is mainly determined by the relaxation of the "slow" particles
and breaking a compact "slow" cluster will be more difficult than breaking 
a less compact cluster. On the other hand, $D$ is predominantly determined
by the diffusion constants of the "fast" moving particles. It will be 
interesting to see whether such an argument works for system where breaking
"slow" particles cluster is even more difficult due to bond formation, for 
example in the "patchy" colloidal system.  

\section{Discussion}  
To conclude we have performed extensive computer simulations of a model two 
dimensional glass forming liquids to understand the breakdown of Stokes-Einstein
relation in these systems. We found that if one randomly pins some fraction 
of particles in the system in their equilibrium positions then degree of 
Stokes-Einstein violation increases dramatically. This enabled us to study
Stokes-Einstein violation in supercooled liquids in a systematic manner and 
also provided us with a model system where exponent related to fractional 
Stokes-Einstein relation can be tuned very precisely. 

Our main findings can be summarized  as follows:
\begin{itemize}
\item	{Breakdown of Stokes-Einstein relation in supercooled liquids get 
enhanced with randomly pinning some faction of particles in the system and
exponent associated with fractional Stokes-Einstein relation also 
systematically increases with increasing pinning concentration.}
\item	{Peak height of $\chi_4(t)$ does not seem to be correlated with 
Stokes-Einstein breakdown in disagreement with previously reported results.}
\item	{The exponent $\beta$ related to stretched exponential decay of the two
point density-density correlation function clearly shows strong correlation
with the Stokes-Einstein breakdown in complete agreement with previous studies.}
\item{Appropriately defined "slow" moving particles over $\alpha$-relaxation
timescale seem to form clusters with decreasing temperature and at round
the temperature where Stokes-Einstein relation starts to violate, the cluster
of "slow" moving particles percolates the whole system size. This indicates a 
possible deep connection between percolation transition of the "slow" particles 
and Stokes-Einstein breakdown.}
\item{Mean cluster sizes of "slow" particles in the system grows with decreasing
temperature and the growth is much stronger for system with larger pinning
concentration.}
\item{The clusters of "slow" particles are fractal like objects with fractal
dimensions $D_f$ in the range $1.5$ to $1.8$ and fractal dimension increases
systematically with increasing pinning concentration. Thus there is a positive
correlation between fractal dimensions of these clusters with the exponent 
related to fractional Stokes-Einstein relation. It will be nice to further
understand this correlation in other model glass forming liquids too.}
\end{itemize} 

The results reported in this study is only for a model two dimensional system, 
so it will be interesting to see whether these results hold for three 
dimensional model system also. It is important to mention here that there are 
recent studies which seem to suggest a qualitative difference in glass 
transition in two and three dimensions \cite{szamelNatComm}. 
Although we don't expect the results reported here to be qualitatively
very different even for three dimensional systems but a detailed study 
is certainly necessary.

\section{acknowledgment}
We would like to thank Kunimasa Miyazaki and J\"urgen Horbach for many 
useful discussions. Kallol Paul and Indrajit Tah are acknowledged for 
critical reading of the manuscript and discussions.

\end{document}